\documentstyle[referee]{mn}
\input psfig.tex
\def\lsim{\lower.5ex\hbox{$\; \buildrel < \over \sim \;$}}
\def\gsim{\lower.5ex\hbox{$\; \buildrel > \over \sim \;$}}
\def \simeq{\lower.3ex\hbox{$\; \buildrel \sim \over - \;$}}
\def\ch{\lower-0.55ex\hbox{--}\kern-0.55em{\lower0.15ex\hbox{$h$}}}
\def\lh{\lower-0.55ex\hbox{--}\kern-0.55em{\lower0.15ex\hbox{$\lambda$}}}
\newif\ifAMStwofonts
\ifoldfss
  \ifCUPmtlplainloaded \else
    \NewTextAlphabet{textbfit} {cmbxti10} {}
    \NewTextAlphabet{textbfss} {cmssbx10} {}
    \NewMathAlphabet{mathbfit} {cmbxti10} {} 
    \NewMathAlphabet{mathbfss} {cmssbx10} {} 
  \fi
  \ifAMStwofonts
    \ifCUPmtlplainloaded \else
      \NewSymbolFont{upmath} {eurm10}
      \NewSymbolFont{AMSa} {msam10}
      \NewMathSymbol{\upi}     {0}{upmath}{19}
      \NewMathSymbol{\umu}     {0}{upmath}{16}
      \NewMathSymbol{\upartial}{0}{upmath}{40}
      \NewMathSymbol{\leqslant}{3}{AMSa}{36}
      \NewMathSymbol{\geqslant}{3}{AMSa}{3E}

      \let\leq=\leqslant 
       
    \fi
  \fi
\fi 
\ifnfssone
  \newmathalphabet{\mathit}
  \addtoversion{normal}{\mathit}{cmr}{m}{it}
  \addtoversion{bold}{\mathit}{cmr}{bx}{it}
  \newmathalphabet{\mathbfit} 
  \addtoversion{normal}{\mathbfit}{cmr}{bx}{it}
  \addtoversion{bold}{\mathbfit}{cmr}{bx}{it}
  \newmathalphabet{\mathbfss} 
  \addtoversion{normal}{\mathbfss}{cmss}{bx}{n}
  \addtoversion{bold}{\mathbfss}{cmss}{bx}{n}
  \ifAMStwofonts
    \ifCUPmtlplainloaded \else
      \UseAMStwoboldmath
      \makeatletter
      \new@mathgroup\upmath@group
      \define@mathgroup\mv@normal\upmath@group{eur}{m}{n}
      \define@mathgroup\mv@bold\upmath@group{eur}{b}{n}
      \edef\UPM{\hexnumber\upmath@group}
      \new@mathgroup\amsa@group
      \define@mathgroup\mv@normal\amsa@group{msa}{m}{n}
      \define@mathgroup\mv@bold\amsa@group{msa}{m}{n}
      \edef\AMSa{\hexnumber\amsa@group}
      \makeatother
      \mathchardef\upi="0\UPM19
      \mathchardef\umu="0\UPM16
      \mathchardef\upartial="0\UPM40
      \mathchardef\leqslant="3\AMSa36
      \mathchardef\geqslant="3\AMSa3E

      \let\leq=\leqslant 

    \fi
  \fi
\fi 

\ifnfsstwo
  \DeclareMathAlphabet{\mathbfit}{OT1}{cmr}{bx}{it}
  \SetMathAlphabet\mathbfit{bold}{OT1}{cmr}{bx}{it}
  \DeclareMathAlphabet{\mathbfss}{OT1}{cmss}{bx}{n}
  \SetMathAlphabet\mathbfss{bold}{OT1}{cmss}{bx}{n}
  \ifAMStwofonts
    \ifCUPmtlplainloaded \else
      \DeclareSymbolFont{UPM}{U}{eur}{m}{n}
      \SetSymbolFont{UPM}{bold}{U}{eur}{b}{n}
      \DeclareSymbolFont{AMSa}{U}{msa}{m}{n}
      \DeclareMathSymbol{\upi}{0}{UPM}{"19}
      \DeclareMathSymbol{\umu}{0}{UPM}{"16}
      \DeclareMathSymbol{\upartial}{0}{UPM}{"40}
      \DeclareMathSymbol{\leqslant}{3}{AMSa}{"36}
      \DeclareMathSymbol{\geqslant}{3}{AMSa}{"3E}

      \let\leq=\leqslant 

    \fi
  \fi
\fi 

\ifCUPmtlplainloaded \else
  \ifAMStwofonts \else 
    \def\upi{\pi}
    \def\umu{\mu}
    \def\upartial{\partial}
  \fi
\fi

\title{Radiatively Driven Plasma Jets around Compact Objects}
\author[Indranil Chattopadhyay and Sandip K. Chakrabarti]
       {Indranil Chattopadhyay$^1$ Sandip K. Chakrabarti$^{1,2}$\\
$^1$ S.N. Bose National Centre for Basic Sciences,\\
JD-Block, Sector III, Salt Lake, Kolkata 700098, India;\\
$^2$ Centre for Space Physics, P61, Southend Garden, Kolkata 700084, India\\
indra@bose.res.in, chakraba@bose.res.in}

\date{Accepted .
      Received ;
      in original form }
\pubyear{}
\begin{document}

\maketitle

\begin{abstract}

Matter accreting onto black holes may develop shocks due to 
the centrifugal barrier. A part of inflowing matter in the post-shock 
flow is deflected along the axis in the form of jets. Post-shock flow  
which behaves like a Compton cloud has `hot' electrons emiting high 
energy photons. We study the effect of these `hot' photons 
on the outflowing matter. Radiation from this region could accelerate 
the outflowing matter but radiation pressure should also slow it down. 
We show that the radiation drag restricts the flow from attaining a 
very high velocity. We introduce the concept of an `equilibrium velocity' 
($v_{eq} \sim 0.5c$) which sets the upper limit of the terminal velocity 
achieved by a cold plasma due to radiation deposition force 
in the absence of gravity. If the injection energy 
is $E_{in}$, then we find that the terminal velocity $v_\infty$ 
satisfies a relation $v_\infty^2 \lsim v_{eq}^2 + 2 E_{in}$.

\end{abstract}


\section{Introduction}

When matter with some angular momentum is pulled in by the
central gravitating object, it spirals in
to form a temporary depository of matter called the accretion disc.
As the accreting matter comes closer to the compact object, 
in much of the parameter space, the centrifugal force becomes 
comparable to the gravitational pull slowing down the flow 
considerably. The flow may suffer a shock in a thin region
(Chakrabarti, 1989; hereafter C89), where the Mach number 
of the flow jumps discontinuously
from supersonic to subsonic value. Entropy is also
generated at the shock. The region in which the flow 
slows down may be extended if the shock conditions 
are not satisfied. This hot, slowed down region puffs up
in the form of a torus (hereafter, CENBOL ${\equiv}$
CENtrifugal pressure dominated BOundary Layer).
This disc solution which includes the advection term may be called 
the advective accretion disc.
Chakrabarti and Titarchuk (1995, hereafter CT95)
pointed out that radiation from this region is responsible for the
characteristic hard and soft states of the black hole candidates.
Similarly, Chakrabarti and co-workers (Chakrabarti, 1998; Chakrabarti, 1999;
Das and Chakrabarti, 1999)
pointed out that the same region may also generate the outflow.
At the shock, a part of the hot accreting matter bounces off this
effective boundary layer of the compact object, and is ejected
along the axis of symmetry as outflows or jets (Das and Chakrabarti, 1999).
Initially, the flow 
should  be subsonic,
since matter comes out with almost zero velocity in a very hot environment.
In this region, the subsonic outflowing matter is
continuously bombarded by hot photons and hence apart from
usual thermal acceleration, radiative acceleration should also be important.

Interaction of radiation and matter in the context 
of black hole astrophysics was investigated as early
as 1974 by Wickramasinghe, where he studied the radiation pressure driven
mass loss from the outer most region of an accretion disc.
Icke (1980) studied the effect of radiative acceleration
of gas flow above a Sakura-Sunyaev Keplerian disc (1973; see also 
Bisnovatyi-Kogan and Blinnikov, 1977). But the
effect of radiation drag on the gas flow was ignored.
Sikora and Wilson (1981) showed that even if the radiation is collimated by
 geometrically thick discs (Lynden-
Bell, 1978; Abramowicz and Piran, 1980), radiation drag is important for 
astrophysical jets.
Piran (1982) while calculating the radiative acceleration of outflows
about the rotation axis of thick accretion discs, found out that in order
to 
accelerate outflows to ${\Gamma}>1.5$ (where ${\Gamma}$ is the
bulk Lorentz factor), the funnels must be short and steep, but such funnels
are found to be unstable.
Sol et. al. (1989) proposed a two flow model for jets, one consists of
relativistic particles (electrons and positrons) and of relativstic
Lorentz factor the other being normal, mildly realitivistic plasma.
Melia and K{\"o}nigl (1989), on the other hand, considered ultra relativistic
jets around super-massive black holes. These jets are assumed to be 
accelerated to super-relativistic
Lorentz factors by hydrodynamic and electromagnetic processes
close to the black hole and are Compton dragged to ${\Gamma}_{\infty}
{\leq}10$.
Icke (1989) considered blobby jets about the axis of symmetry of
thin discs and he obtained the `magic speed'
of $v_{m}=0.451c$ where $c$ is the velocity of light. 
Fukue (1996) went one step 
ahead and found out that outflowing-rotating matter, away from the axis of 
symmetry, achieves terminal speed less than what Icke found previously.
Sikora et. al. (1996) found out that the net radiative acceleration of AGN
jets,
vanishes for values of the ${\Gamma}_{eq}{\leq}4$. 
Recently Fukue and his collaborators (Watarai et. al. 1999; Hirai and
Fukue 2001, Fukue et. al. 2001) systematically studied radiative 
acceleration and collimation of
winds from a disc, which is a combination of outer luminous slim disc and inner
advection dominated region. 
In the present paper, we investigate the issue of the radiative acceleration of
outflows by radiation coming out from advective accretion discs (C89; CT95;
Chakrabarti, 1990a; Chakrabarti, 1996).
In the papers mentioned above, in order to mimic cold plasma the 
pressure gradient term was altogather neglected. However, we include 
the pressure gradient term as well.
A number of workers,
such as Castor (1972), Hsieh and Spiegel (1976),
Blandford and
Payne (1981), Mihalas and Mihalas (1984, here after MM84), Fukue et al (1985), 
Kato et. al. (1998) have investigated the equations of
photo-hydrodynamics. 
The equations of motion for non-rotating outflow is taken from MM84.
We solve the governing equations of matter and radiation by the
so-called `sonic point analysis' (Chakrabarti, 1990b; here after C90b).
The radiation energy density and flux along the axis of symmetry
is calculated following the treatment of Chattopadhyay and Chakrabarti (2000).

In \S 2.1, we present the model assumptions. In \S 2.2, we present
the equations of motion of the cold outflowing plasma, and in the rest 
of the Section we discuss the method of solution. In \S 3, we present the
solutions, where we find that more radiation results in more acceleration,
though radiation drag has a limiting effect on the terminal velocity
achieved. Finally, in \S 4, we draw our conclusions.  

\begin {figure}
\vbox{
\vskip -0.0cm
\hskip 0.0cm
\centerline{
\psfig{figure=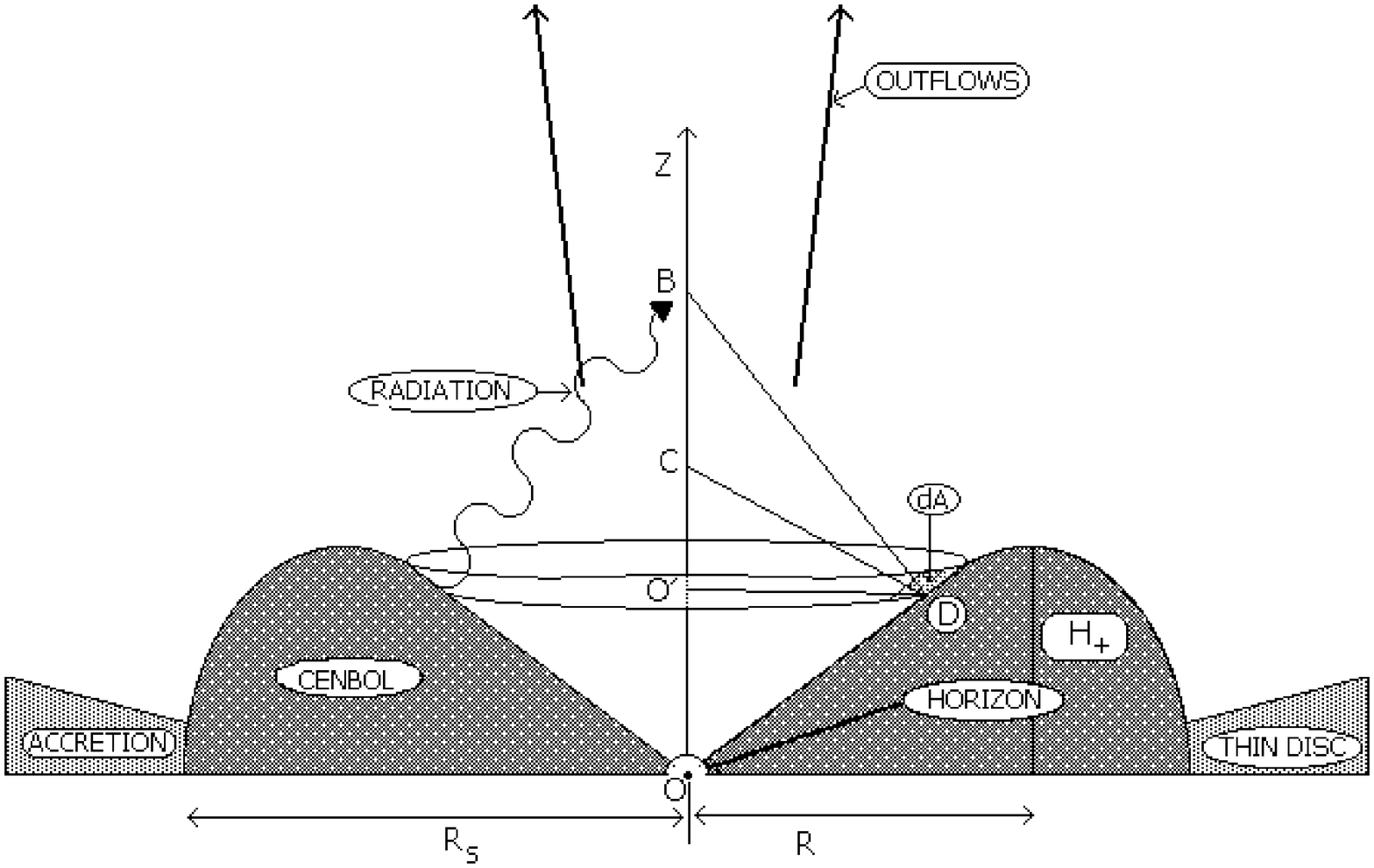,height=10truecm,width=16truecm}}}
\end{figure}
\begin{figure}
\vspace{-0.2cm}
\caption[] {Schematic diagram of CENBOL
and disk/outflow geometry. $D$ is the source point, $B$ is the field point.
$O$ is the position
of compact object. The local unit normal $\hat{n}$ is along
$DC$. $OO^{'}=Z^{'}$, $O^{'}D=X$.
}
\end{figure}

\section{\bf Assumptions, Governing Equations and the
Method of Solution }

\subsection{ Assumptions}

We assume that the outflows are non-viscous and non-rotating. As the
astrophysical jets or outflows are observed to be extremely collimated
(Bridle and Perley, 1984),
the flow geometry is considered to be thin and conical. Thus the transverse
structure of the jet is ignored which means all the flow variables will
be calculated on the axis of symmetry and will be assumed not to
vary along the transverse direction.

In this paper we are not considering the origin of the outflowing matter
self-
consistently ---
instead, we have extracted its essential features while doing
radiative-hydrodynamics. As outflows originate
from the post-shock region in accretion, its initial velocity has to be
necessarily
subsonic. In the present analysis, this becomes the inner boundary
condition of the outflow.
The only connection with
accretion considered here is the computation of the radiation flux,
the energy density and
the size of the base of the jet (i.e., CENBOL).

We are basically interested to investigate the issue of radiative
acceleration of outflows by momentum deposition of photons to the electrons.
In Compton scattering, the transfer
of energy per scattering (Rybicki and Lightman, 1979) is given by,
$$
{\Delta}{\it e}=\frac{{\it e}}{m_ec^2}(4{k_B}{T_e}-{\it e}),
\eqno{(1)}
$$

\noindent where, ${\Delta}{\it e}=$ net change of photon energy,
${\it e}$ is the
photon energy, $k_B$ is the Boltzmann constant and $T_e$ is the
electron temperature. It is clear that
if ${\it e}>4{k_B}{T_e}$ then energy is transferred from photons to electrons.
Hence for radiative acceleration it is necessary to consider
flows to be `cooler' compared to the radiation though this by itself
does not ensure that the energy thus supplied can be converted
to bulk kinetic energy of the outflow.  \\

\subsection{Governing Equations}

Equations of motion for matter under the action of radiation has been 
extensively studied in the papers mentioned in \S 1. We use the equation of
motion given by MM84 for radial flow of grey medium.
We have assumed the radiation pressure to be equal to the energy density
and the Eddington factor is of the order of one (see, MM84). 
To take care of the strong gravity around a static compact object
we introduce the pseudo-Newtonian
gravitational potential (Paczy\'nski and Wiita, 1980).
Equations governing the steady state flow, in the pseudo-Newtonian
limit become, \\
\noindent {\it Momentum Balance Equation:}

$$
v\frac{dv}{dz}+\frac{1}{\rho}\frac{dP}{dz}+\frac{1}{2(z-1)^2}=
{\cal D}_z-2 v {\cal E}_z ,
\eqno{(2)}
$$

\noindent {\it Baryon Number Conservation in the Outflow Equation:}

$$
{\dot M}={\Theta}_{out}{\rho}vz^2 ,
\eqno{(3)}
$$
\noindent where
$$
{\Theta}_{out}=2{\pi}{{tan}^2}{\theta}_0 ,
$$
\noindent and {\it Entropy Generation Equation:}

$$
v(\frac{d{\epsilon}}{dz}-\frac{P}{{\rho}^2}\frac{d{\rho}}{dz})=
Q_{+}-Q_{-}.
\eqno{(4)}
$$

\noindent The variables $v$, $\rho$ and $P$ in the above equations are,
velocity, density and isotropic gas pressure respectively of the flow.
${\theta}_0$ is the opening angle of the conical outflow.
In the above equations, ${\cal D}_z=\frac{{\sigma}_T}{m_p}F_z$,
where $F_z$ is
the z-component of flux of radiation on the axis of symmetry and ${\cal E}_z=
\frac{{\sigma}_T}{m_p}{\it E}$, where ${\it E}$ is the radiation energy density
on the axis of symmetry. One must note, ${\cal D}_z$ is the radiative
acceleration term and is abbreviated as RAMOD (RAdiative MOmentum
Deposition force) and the second term in RHS of Eq. (2) is the 
radiation drag term. ${\cal D}_z$ and ${\cal E}_z$ is calculated following
the prescription of Chattopadhyay and Chakrabarti (2000), which is 
given in \S 2.3, for the sake of completeness.
The third term on the LHS of Eq. (2)
is the gravity with `Paczy\'nski-Wiita' (psuedo-Newtonian) potential.
The first term ($Q_+$) in LHS of Eq. (4) is the absorption term 
or any other heating term, which is considered 
to be zero. The second one is the cooling (bremsstrahlung, synchrotron etc)
term. We neglect cooling terms. Later in this paper, we shall come back to 
them.
$\epsilon$ in LHS of Eq. (4) is the specific internal energy
term and ${\epsilon}=\frac{P}{({\gamma}-1){\rho}}$.
As jets are extremely collimated, they are assumed to be
in a narrow cone about the axis of symmetry (z-axis), moving radially
outward, we choose
${\theta}_0 {\approx} 10^{o}{\approx}$ opening angle of the bipolar outflow,
considered to be constant here.
The gas is assumed to obey an ideal equation of state,
$$
P=\frac{{\rho}k_BT}{{\mu}m_p}.
\eqno{(5)}
$$
Here $T$ is the
temperature, $m_p$ is the
proton mass and  ${\mu}=0.5$.
\noindent We define the adaiabatic sound speed $a$ as,
$$
a^2=\frac{\gamma P}{\rho}.
\eqno{(6)}
$$

Units of mass, length and time are chosen to be $M_B$ (Mass of the compact
object), $2\frac{GM_B}{c^2}$ (Schwarzschild radius;
G=Universal Gravitational constant) and $2\frac{GM_B}{c^3}$ respectively,
also $c=1$.

\subsection{Calculation of $D_z$}

Figure 1 shows the schematic diagram of a disc/jet geometry.
The flux of radiation at field point $B(z)$, from differential
area $dA$ around a source point $D(X, {\phi}, Z^{'})$ is given by:
$$
d{\bf F}=I{\cos}{\angle (CDB)}d{\Omega}{\frac {\bf DB}{|{\bf DB}|}} ,
\eqno{(7)}
$$
where $I$ is the frequency averaged radiation intensity  on the surface
and $d{\Omega}$ is the solid angle subtended by the differential area $dA$ at
$B$.
The net momentum transferred (in dimensionless units) at B,
due to the annular area on the cone,
between $Z^{'}$ and $Z^{'}+dZ^{'}$, along $z$ axis is given by:
$$
d{\cal D}_{z}={\frac{2{\pi}I{\sigma_{T}}}{m_{p}}}{\tan}^{2}{\theta}
{\frac{z(z-Z^{'})Z^{'}}{[(z-Z^{'})^{2}+(Z^{'}{\tan}{\theta})^{2}]^{2}}}dZ^{'}
,
\eqno{(8)}
$$
where, ${\sigma}_{T}$ is the Thomson scattering cross section.
${\theta}$ is the semi-vertical angle of the inner surface of the
CENBOL (see, Fig. 1) which is assumed to remain constant. The
direction
of $d{\cal D}_{z}$ is along $z$ axis, as the transverse component
of the momentum transferred, due to each of the differential area
$dA$, cancels each other as we integrate over ${\phi}$.
\noindent We integrate over $Z^{'}$ to get the entire radiative
contribution
of the CENBOL, and hence ${\cal D}_{z}$ (RAMOD).
\noindent Thus the analytical expression of RAMOD along $z$ is:

$$
{\cal D}_{z}={\cal D}_0[{\cos}2({\Phi}_{1}-{\theta})-{\cos}2({\Phi}_{2}
-{\theta})] ,
\eqno{(9)}
$$

\noindent where
$$
{\Phi}_{1}=tan^{-1}\left [\frac{2(H_{+}-z{\cos}^{2}{\theta})}
{z{\sin}2{\theta}}\right ] ,
$$

\noindent and

$$
{\Phi}_{2}=tan^{-1}\left [\frac{2(H_{-}-z{\cos}^{2}{\theta})}
{z{\sin}2{\theta}}\right ] .
$$
\noindent $H_{+}$ and $H_{-}$ are maximum and minimum height of the CENBOL
from the
equatorial plane, respectively, and
$$
{\cal D}_0={\frac{{\pi}I{\sigma_{T}}}{2m_{p}}} .
\eqno{(10)}
$$
Now, we have to make an estimate of
$I$. We assume that the radiation coming out is the
binding energy of the last stable orbit of accreting
matter. Hence,
$$
I=\frac{L}{{\Omega}_{T}{\cal A}} .
\eqno{(11)}
$$

\noindent $L={\eta}{{\dot M}_{acc}}c^2$, is the amount of rest mass
energy of the accreting matter converted into radiation per unit
time. Here ${\dot M}_{acc}$ is the accretion rate. ${\eta}$
is the convertion ratio and is equal to $0.06$ for the
Schwarzschild metric.
${\Omega}_{T}$ is the solid angle in which the radiation is
coming out locally.
${\cal A}$ is the total area of the inner surface of the
CENBOL. I is assumed to be uniform here.

\subsection{Calculation of ${\cal E}_{z}$ }

In reference to Fig. 1, one can also calculate the radiation energy
density. The radiation energy density at the field point $B(z)$ is
given by:
$$
d{\it E}=\frac{I}{c}{\cos}{\angle (CDB)}d{\Omega}.
\eqno{(12)}
$$
\noindent Integrating over the inner surface of the CENBOL
as in the previous case we have:
$$
{\it E}={\it E}_0 z
\left[-\frac{1}{\{ (z^{'}-z{\cos}^2{\theta})^2+z^2{\sin}^2{\theta}
cos^2{\theta}\}^{1/2} } +\frac{1}{z{\sin}^2{\theta}}{\sin}
\{ {\tan}^{-1}\frac{z^{'}-z{\cos}^2{\theta}}{z{\sin}{\theta}
{\cos}{\theta}} \} \right]_{H_{-}}^{H_{+}} .
$$

\noindent Hence,
$$
{\cal E}_{z}={\cal E}_0 z
\left[-\frac{1}{\{ (z^{'}-z{\cos}^2{\theta})^2+z^2{\sin}^2{\theta}
cos^2{\theta}\}^{1/2} } +\frac{1}{z{\sin}^2{\theta}}{\sin}
\{ {\tan}^{-1}\frac{z^{'}-z{\cos}^2{\theta}}{z{\sin}{\theta}
{\cos}{\theta}} \} \right]_{H_{-}}^{H_{+}} ,
\eqno{(13)}
$$
\noindent where,
$$
{\cal E}_0=2{\pi}{{\sin}^2{\theta}}{\cos}{\theta}\frac{I{\sigma}_T}{m_p} ,
\eqno{(14)}
$$
\noindent and
quantities $[...]_{H_{-}}^{H_{+}}$ refers to $[...]_{H_{+}}-[...]_{H_{-}}$.
We have chosen the shock in the accretion flow to be located
at distance
$r=R_s=10r_g$ (see CT95) where $1r_g=2G{M_B}/{c^2}$ and
the specific angular momentum
of the accretion to be $1.6(2G{M_B}/c)$. This determines the
geometrical structure of the CENBOL. From Fig. 1, we see that $R$
is the position of $H_{+}$ on the equatorial plane of the
accretion disc. And $R_s$ is the maximum radial distance of the
CENBOL. With the above choice of accretion parameters, we have
$H_{+}=4.4r_g$ and $R=6.671r_g$. $H_{-}=1.1r_g$. With these
geometric structure of CENBOL is determined which in turn helps
to calculate ${\cal D}_z$ and ${\cal E}_z$. Though it seems that
geometric structure of CENBOL introduces additional parameters,
but one must remember that these along with the total amount of
radiation, are dependent on the accretion parameters.
So self-consistent accretion-jet solution should determine these
from the accretion parameters (namely, accretion rate and angular
momentum) only.

\subsection{ Method}

{\it Sonic Point Analysis}:
To solve Eqs. (2)-(4) we use the method of sonic point analysis (C90b).
In this paper the outflow is assumed inviscid and fully ionised plasma.
Moreover, only electron scattering is considered. Thus, $Q_{+}=0$ in 
Eq. (4). The cooling terms are not considered for the time being
(we will come back to them in later part of \$ 3), hence $Q_{-}=0$.
Equations (2)-(4) with the help of Eqs. (5) and (6) can be written as;
$$
\frac{dv}{dz}=\frac{\frac{2a^2}{z}-\frac{1}{2(z-1)^2}+{\cal D}_z-2v{\cal E}_z}
{v-\frac{a^2}{v}}
\eqno{(15)}
$$

\noindent and

$$
\frac{da}{dz}=
-\frac{a(\gamma-1)}{z}-\frac{a(\gamma-1)}{2v}\frac{dv}{dz} .
\eqno{(16)}
$$

Since an outflow around compact objects must originate from inflowing
matter, it must
start with a subsonic
velocity. Therefore relativistic outflows which acquire supersonic
terminal
velocity
at large distances from the central object, must be transonic in nature.
Thus, at some point (critical or sonic point), the
denominator
of Eq. (15) must go to zero.  Since the flow must be smooth everywhere,
this implies that at
radial
distance
where the denominator of Eq. (15) goes to zero, the numerator
must go to zero as well. Using this property we have the so-called
{\it critical point conditions} or {\it sonic point conditions}.
The sonic point conditions are :

$$
v_c^2=a_c^2,
\eqno{(17)}
$$

\noindent and

$$
a_c^2-{\cal E}_{z_c}a_c+\frac{z_c}{2}{\cal D}_{z_{c}}-
\frac{z_c}{4(z_c-1)^2}
=0,
\eqno{(18)}
$$

\noindent where $z_c$ is the sonic point. By definition it is the
distance on the axis of symmetry at which the sonic point conditions
are satisfied.
As $z {\rightarrow} z_c$, $\frac{dv}{dz}=\frac{N}{D}
{\rightarrow} \frac{0}{0}$. The gradient of radial velocity
at $z_c$ is found out by l'H\^ospital's rule,
$$
{\left(\frac{dv}{dz}\right)}_c=
{\left[\frac{\frac{dN}{dz}}{\frac{dD}{dz}}\right]_c} .
\eqno{(19)}
$$

\noindent The expression of gradient of sound speed at $z_c$,
is obtained by modifying Eq. (16),

$$
{\left(\frac{da}{dz}\right)}_c=-\frac{a_c(\gamma-1)}{z_c}
-{\frac{(\gamma-1)}{2}}{\left(\frac{dv}{dz}\right)}_c .
\eqno{(20)}
$$

To obtain a complete transonic solution, we supply $z_c$,
${\dot M}_{acc}$
and integrate Eqs. (15) and (16), with the help of Eqs. (17)-(20),
once from $z_c$ inwards and
again from $z_c$ outwards to a large distance.
Once we obtain $v$ and $a$ as
functions of $z$, we have a complete solution of the problem.
Other flow variables like $T$ can be found out from
Eqs. (5) and (6) and $\rho$ from Eq. (3).
To get an estimate of the electron temperature, we can
use the expression ${T_e}=({\frac{m_e}{m_p}})^{1/2}T_p$.
Calculation of $\rho$ and estimation of $T_e$ are used when cooling
mechanisms are considered.
To calculate $\rho$, one has to supply the
value of ${\dot M}$, the only constant of motion.
>From the works of Chakrabarti and his collaborators, it is known that
${\dot M}$ is not a
free parameter, and can be calculated as a function of
${\dot M}_{acc}$ and the compression ratio at the shock
of the accretion disc.
We make an estimation of ${\dot M}$ from ${\dot M}_{acc}$,
following Chakrabarti (1999). \\
\noindent Here the accretion and wind is assumed to be
conical.
According to
Chakrabarti (1999),

$$
\frac{\dot M}{{\dot M}_{acc}}=\frac{{\Theta}_{out}}{{\Theta}_{acc}}
\frac{R_c}{4}{[\frac{{R_c}^2}{R_c-1}]^{3/2}}{\rm exp}(\frac{3}{2}-
\frac{R_c^2}{R_c-1})
\eqno{(21)}
$$

\noindent where, ${\dot M}$ is the mass outflow rate, ${\dot M}_{acc}$
is the mass accretion rate and $R_c$ is the compression ratio at the
shock in accretion. ${\Theta}_{out}$ is the solid angle subtended by
the outflow, ${\Theta}_{acc}$ is the solid angle subtended by the
accreting matter.
Assuming ${\Theta}_{out} {\sim} {\Theta}_{in}$ (for simplicity)
Chakrabarti calculated the mass outflow rate for any generic
value of compression ratio, the latter varying from $1$ (no-shock)
to $7$ (very strong shock). 
Putting $R_c=1$ (no-shock) in Eq. (21) Chakrabarti found out ${\dot M}
=0$ , i.e., there is no outflow if there is no shock in accretion.
The maximum outflow rate calculated
was about $30 {\%}$ of the accretion (for $R_c{\sim}4$). 
In the strong shock limit (for $R_c{\sim}5$)
it was calculated to be less than $10 {\%}$.
In the present work, the mass outflow rate is assumed to be,
$$
{\dot M}=0.01{\dot M}_{Edd},
\eqno{(22)}
$$
\noindent until otherwise stated. Of course, the actual
value of mass outflow rate is necessary only while
considering effects of
various cooling
mechanisms.

\begin{figure}
\vbox{
\vskip -0.0cm
\hskip 0.0cm
\centerline{
\psfig{figure=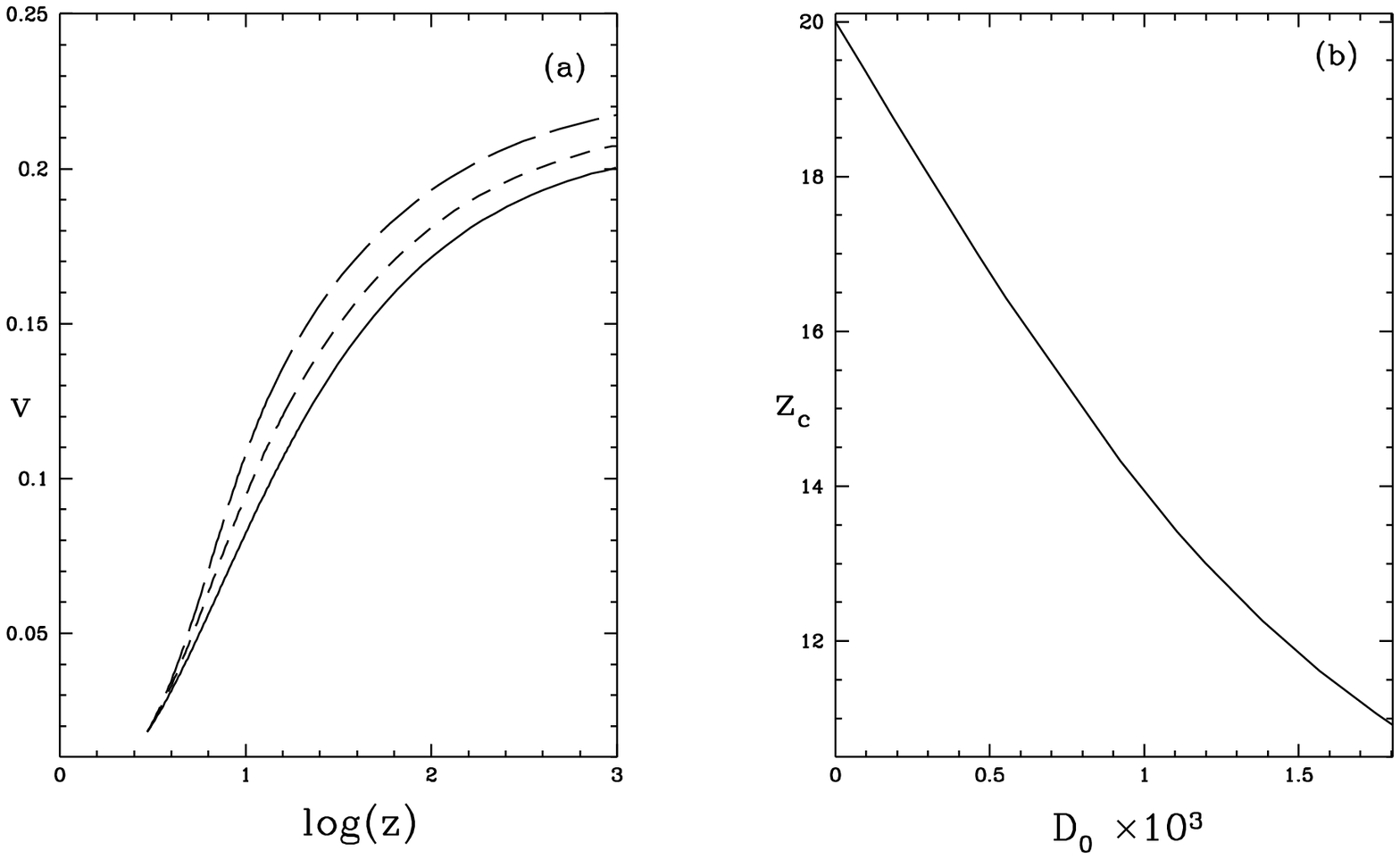,height=14truecm,width=14truecm}}}
\end{figure}
\begin{figure}
\vspace{-5.0cm}
\caption[]
{Solution of outflow variables plotted
with $log(z)$. 
Initial parameters are $z_{in}=2.97$, $v_{in}=0.02$.
(a) Comparison of velocity variation ($v$)
for RAMOD corresponding to ${\dot M}_{acc}=12{\dot M}_{Edd}$
(long-dashed), ${\dot M}_{acc}=6{\dot M}_{Edd}$ (short-dashed) and
Bondi-type (solid) outflow.
(b) Variation of sonic point $z_c$ with
RAMOD (${\cal D}_0$ is the space independent part of ${\cal D}_r$). }
\end{figure}

\section{Results}

In view of RHS of Eq. (4) to be zero, Eq. (2) can be integrated and can be 
written in the form,
$$
\frac{1}{2}{v(z)}^{2}+n{a(z)}^{2}-\frac{1}{2(z-1)}=
\frac{1}{2}{v_{in}}^{2}+n{a_{in}}^{2}-\frac{1}{2(z_{in}-1)}+
{\int_{z_{in}}^{z}}({\cal D}_{z^{'}}-2v{\cal E}_{z^{'}})dz^{'}
\eqno{(23a)}
$$
or,
$$
E(z)=E_{in}+{\int_{z_{in}}^{z}}({\cal D}_{z^{'}}-2v{\cal E}_{z^{'}})dz^{'}
\eqno{(23b)}
$$

Equation (23) shows that, in absence of radiation, non-rotating outflows 
are accelerated 
by thermal pressure only and such outflows are called Bondi-type (1952)
outflows.
>From the above equation, it is quite clear that in order to 
accelerate outflows 
{\it significantly} by radiation, the work done by
radiation forces, onto the flow has to be 
{\it comparable} to the initial energy of the outflow. 
We will come back to this
equation to understand quite simply various phenomenon. 

\begin {figure}
\vbox{
\vskip 0.0cm
\hskip 0.0cm
\centerline{
\psfig{figure=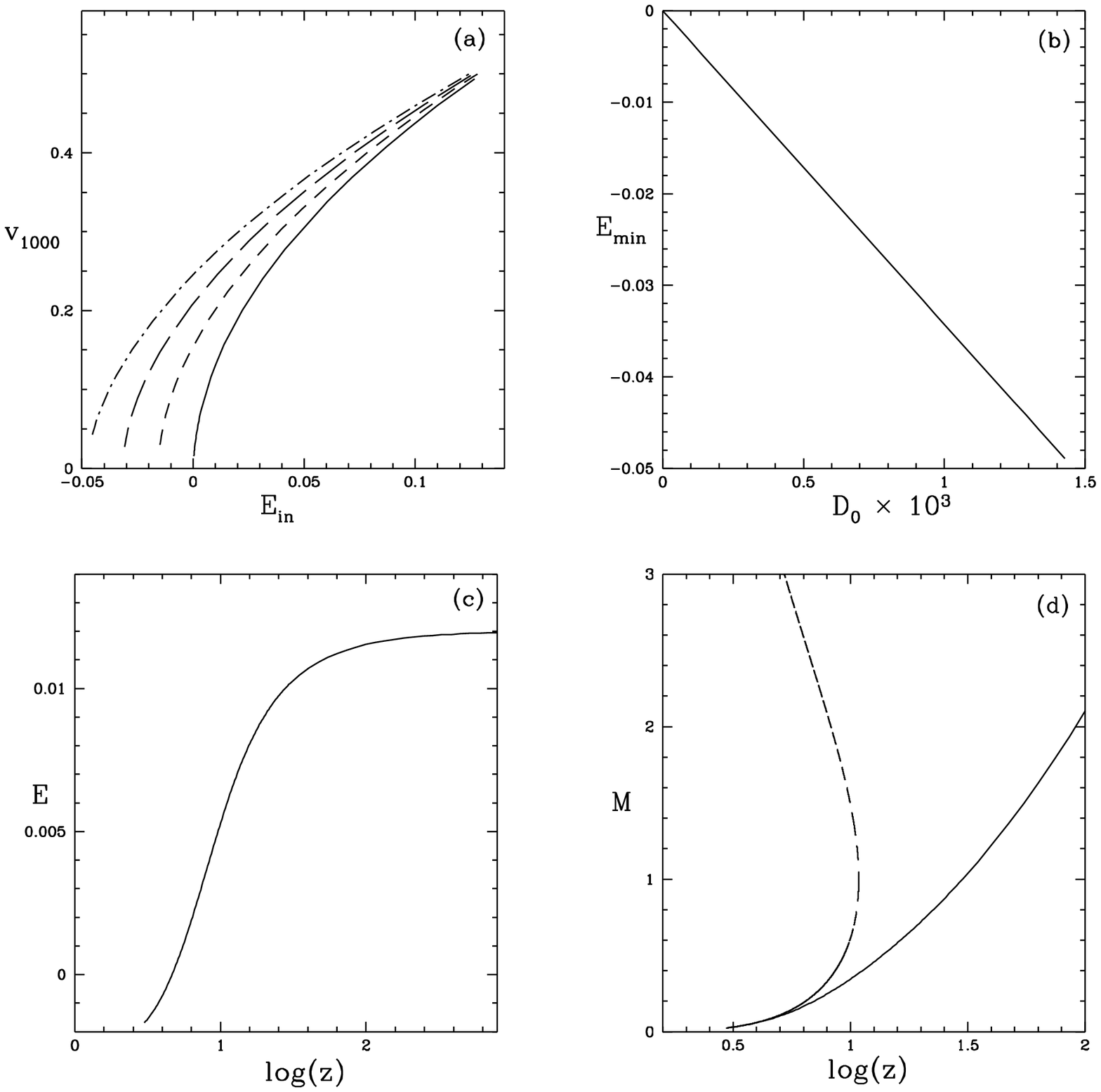,height=14truecm,width=14truecm}}}
\end{figure}
\begin{figure}
\vspace{-0.0cm}
\caption[] { 
(a) Comparison of ($v_{1000}$) with initial
energy (${\rm E}_{in}$) of the flow. Various curves represent
flows acted on by RAMOD corresponding to ${\dot M}_{acc}=15{\dot M}_{Edd}$
(dash-dot),
${\dot M}_{acc}=10{\dot M}_{Edd}$
(long-dashed), ${\dot M}_{acc}=5{\dot M}_{Edd}$ (short-dashed) and
Bondi-type (solid) outflow. (b) Variation of the critical specific 
energy of the flow ($E_{min}$) with ${\cal D}_0$.
(c) Variation of sp. energy $E$ of the flow as a function of distance.
RAMOD corresponds to ${\dot M}_{acc}=5{\dot M}_{Edd}$. Initial parameter is
${\rm E}_{in}=-0.002$ at $z_{in}=3$. (d) Comparison of the variation
of Mach number ($M$) with $log(z)$. The solid curve represents the outflow
acted on by RAMOD corresponding to ${\dot M}_{acc}=5{\dot M}_{Edd}$. The
dashed curve represents the outflow driven only by the thermal energy. Initial
parameter is, ${\rm E}_{in}=-0.002$ at $z_{in}=3$ for both the cases. }
\end{figure}

Let us now focus on the solution topologies of the outflow which is
accelerated by the
radiative momentum
deposition. Figure 2a, shows the variation of bulk velocity ($v$) along the
axis of symmetry with $log(z)$. Various curves denote flows acted on by
radiation corresponding to ${\dot M}_{acc}=12{\dot M}_{Edd}$ (long-dashed),
${\dot M}_{acc}=6{\dot M}_{Edd}$
(short-dashed) and the solid line denotes
Bondi-type outflow. The initial
parameters are the same in all the curves $---$ $v_{in}=0.02$ (in units of $c$)
at $z_{in}=2.97$ (in units of $2GM/{c^2}$).
We see that the outflows are accelerated 
by radiative momentum deposition. As outflows with the same input
parameters are accelerated to higher and higher terminal velocities,
correspondingly
sonic point will also come closer. Figure 2b, exhibits this phenomena where
the sonic point ($z_c$) is plotted with non-spatial part of RAMOD 
[${\cal D}_0$;
see, Eq. (10)]. Input parameters are the same as that of Fig. 2(a). 

\begin {figure}
\vbox{
\vskip -0.0cm
\hskip 0.0cm
\centerline{
\psfig{figure=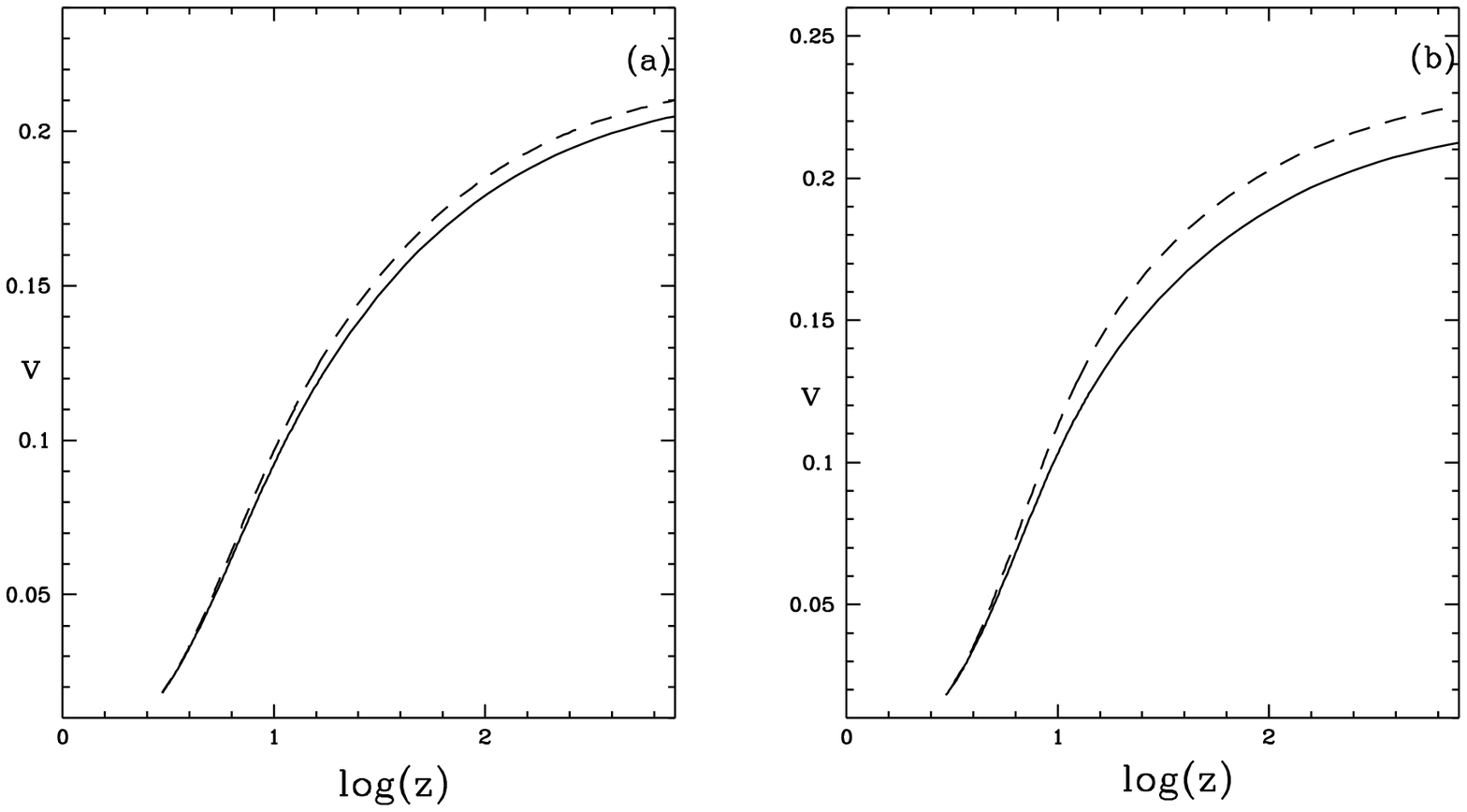,height=14truecm,width=14truecm}}}
\end{figure}
\begin{figure}
\vspace{-3.0cm}
\caption[] {Comparison of velocity variation of outflows with (solid)
and without (short-dashed) radiation drag. RAMOD corresponds to 
(a) ${\dot M}_{acc}=5{\dot M}_{Edd}$
and (b) ${\dot M}_{acc}
=10{\dot M}_{Edd}$; $z_{in}=2.96$, $ v_{in}=0.02$.} 

\end{figure}

Figure 3(a), shows the variation of $v_{1000}$
($v$ at $z=1000$)
with initial energy (${\rm E}_{in}$) of the flow for outflows
acted on by RAMOD
proportional to ${\dot M}_{acc}=15{\dot M}_{Edd}$ (dash-dot),
${\dot M}_{acc}=10{\dot M}_{Edd}$ (long-dashed),
${\dot M}_{acc}=5{\dot M}_{Edd}$ (short
-dashed) and the Bondi-type outflow is denoted by the
solid line. We see that though $v_{1000}$ achieves a higher value, 
it is almost
independent of RAMOD when ${\rm E}_{in}$ is high. 
For lower value of ${\rm E}_{in}$, the radiative momentum
deposition plays more significant role
and even outflows which are initially bound are pushed
to infinity as transonic outflows. 
>From Eq. (23b) we see that if $E_{in}$ is significantly greater than 
radiative work done then clearly $v$ at large $z$  will have a weak 
dependence on radiation. More over, we also see
that, if LHS of Eq. (23b) is less than zero then there is no outflow.
Since outflows originate close to the black hole, for cold plasma
$E_{in}$ may be negative. In that case the wind will blow if the 
second term in RHS of
Eq. (23a) is greater than $E_{in}$. Thus the critical specific initial energy
$E_{min}$ for which 
wind do not blow, will be obtained when these two terms are equal. Figure 3(b),
shows $E_{min}$ as a function of ${\cal D}_0$. Outflows with $E_{in}$
on this curve and bellow will not reach a large distance. 

\begin {figure}
\vbox{
\vskip 1.5cm
\hskip -0.0cm
\centerline{
\psfig{figure=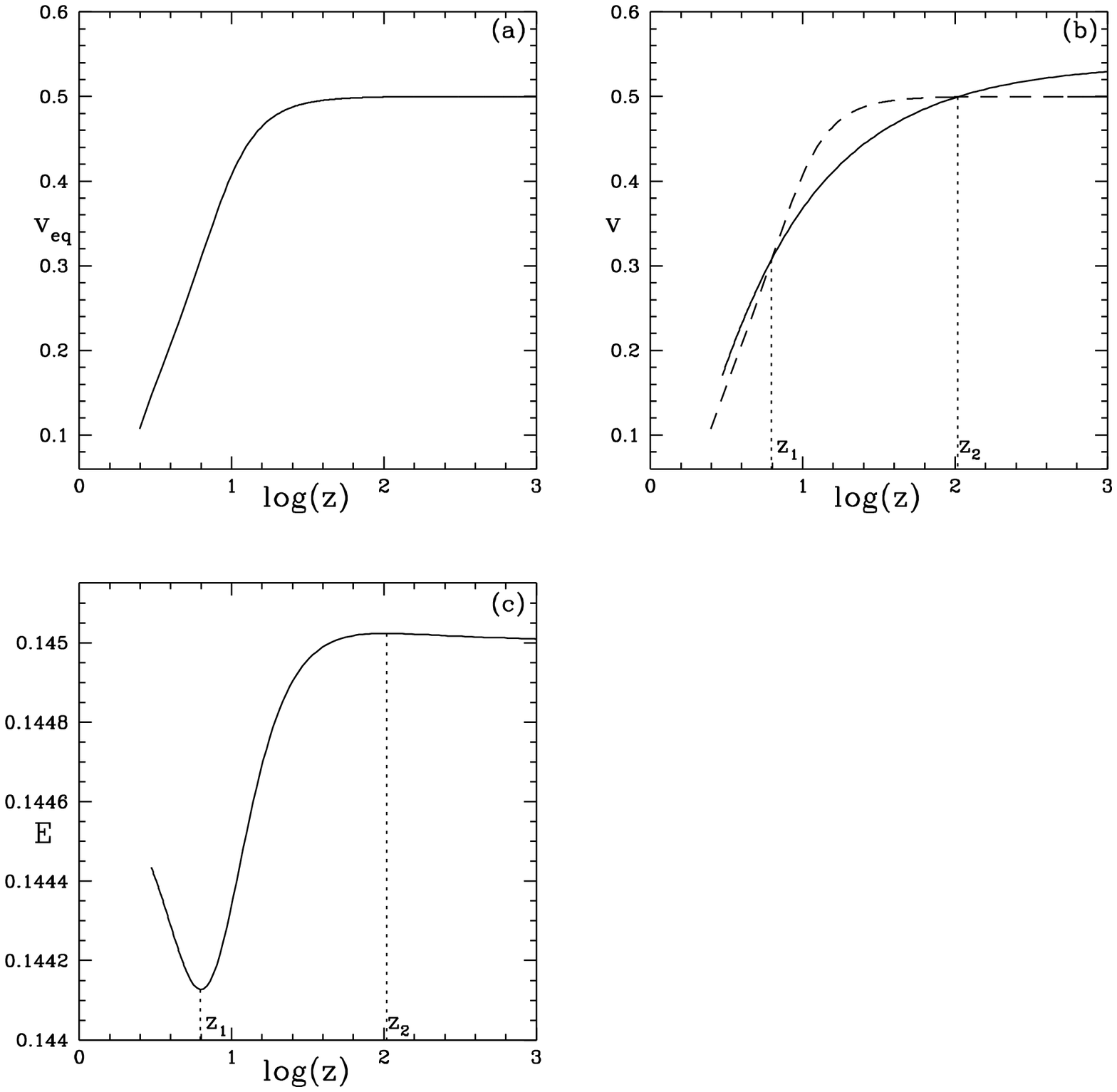,height=14truecm,width=14truecm}}}
\end{figure}
\begin{figure}
\vspace{-0.2cm}
\caption[] {(a) Variation of $v_{eq}$ as a function of $z$
(b) Comparison of variation of bulk velocity $v$ (solid) with $v_{eq}$ 
(short-dashed) as a function of $z$. $v_{in}=0.17$ at $z_{in}=2.96$.
$v_{eq}(z_{in})<v_{in}$. The outflow topology (solid) is acted on by RAMOD
corresponding to 
${\dot M}_{acc}=5{\dot M}_{Edd}$. (c) Comparison of $E$ of the 
outflow in the previous case. }

\end{figure}

Figure 3(c) shows the variation of specific
energy of the outflow which was initially bound, energetically speaking.
Initial
parameters are ${\rm E}_{in}=-0.002$ (in units of $c^2$) at $z_{in}=3$.
Consequently, the Mach number ($M=v/a$) variation of the same solution 
(solid) is
compared with only the thermally driven outflow (short-dashed), starting with
the same initial condition we find no transonic solution at all, as is
exhibited in Fig. 3(d).

We now investigate the role of radiative deceleration
or the radiation-drag term i.e., the third term in the RHS of
the momentum balance equation. The radiation drag term is proportional
to both $v$ and ${\cal E}_z$. i.e., as $v$ increases radiation drag also 
increases. But if the radiation energy density increases, the drag term will
increase too. This phenomenon is exhibited in Fig. 4.  
Figure 4(a),
shows the comparison of velocity ($v$) variation of outflows with (solid)
and without (short-dashed) radiation drag. RAMOD corresponds to
${\dot M}_{acc}=
5{\dot M}_
{Edd}$ was chosen.
Initial parameters are $z_{in}=2.96$,
$v_{in}=0.02$.
Figure 4(b), shows the comparison of velocity ($v$) variation of outflows 
with (solid)
and without (short-dashed) radiation drag for RAMOD corresponding to
${\dot M}_{acc}=10{\dot M}_{Edd}$. 
Increasing the radiation one would have expected more acceleration
and hence more $v_{1000}$ at large $z$. But as this also increases
the radiative energy density the drag term increases and results in 
marginally higher $v_{1000}$.
As radiative deceleration is proportional to both $v$ and radiative energy
density, the actual deceleration will depend on competition of both of
these terms. Outflows are generated close to the black hole,
radiation energy density
due to the inner part of the disc is large and initial velocity of the
outflow is small. Hence, initially radiation drag is not
strong enough
to stop the outflow from forming. As outflows achieve
supersonic velocity far away from the black hole, radiation energy
density falls off, thus radiation drag once again is not strong enough to
make the outflow decelerate or for that matter
to fall back on the disc. 
Now the question is what is the maximum velocity allowed before deceleration 
sets in.
If we equate the two terms in RHS of Eq. (2), we get the equilibrium
speed above which there would be deceleration.
Thus,
$$
v_{eq}=\frac{{\cal D}_z}{2{\cal E}_z}
\eqno{(24)}
$$

Thus we see that the maximum velocity beyond which deceleration sets in
depends on both ${\cal D}_z$ and ${\cal E}_z$. Inside the funnel shaped
region ${\cal E}_z$ is large, but ${\cal D}_z$ is small, as flux of radiation
along $z$ is smaller. As outflowing matter leaves the funnel ${\cal D}_z$
increases compared to ${\cal E}_z$ and consequently $v_{eq}$ also increases. 
The dashed curve in Fig. 5(a) 
shows the variation
of $v_{eq}$ as a function of $z$. As $z{\rightarrow}$ large, 
$v_{eq}{\to} 0.5$. This is what Icke(1989) called as the `magic
speed', which for a thin disc was found to be equal to $0.451$. 
Important thing to note is that though magic speed or in our
parlance $v_{eq}$ at large $z$, does not depend on the magnitude 
of radiation intensity 
from the CENBOL, but actual deceleration will. 
In Fig. 5(b), matter is injected with $v_{in}$ larger than $v_{eq}(z_{in})$.
In the region $z<z_1$, $v>v_{eq}$ so
there is a net deceleration and the flow velocity decreases than what it should
be. At $z=z_{1}$ $v(z_1)=v_{eq}(z_1)$ which means that
the radiative force is zero. For $z_2>z>z_{1}$,
$v<v_{eq}$, radiation again accelerates the flow. At $z>z_2$, again
$v>v_{eq}$, the flow again decelerates but as at large $z$, net
radiative deceleration is small, the effect is marginal. If one 
looks at the variation of $E$ of such a flow, the effect is more 
understandable. In Fig. 5(c), specific energy of the flow [same as
Fig. 5(b)] is plotted with $z$. In the region $z<z_1$, there is net
deceleration so $E$ decreases. At $z>z_1$, there is acceleration, hence
$E$ increases. Again at $z>z_2$, $v>v_{eq}$ which means there is 
radiative deceleration but as the magnitude of deceleration is small,
the effect is marginal.
Note however, $v_{eq}$ is not the terminal velocity.
For $z {\rightarrow} {\infty}$, $E(z){\rightarrow}{\frac{1}{2}}v_{\infty}^{2}$.
From Eq. (23) we get,
$$
\frac{1}{2}v_{\infty}^{2}=E_{in}+{\int_{z_{in}}^{\infty}}({\cal D}_{z^{'}}
-2v{\cal E}_{z^{'}})dz^{'} .
\eqno{(23c)}
$$
The above equation tells us that $v_{\infty}$ will depend on both
the net radiative work done on to the flow and also the initial
energy of the flow. 

Apart from looking at acceleration phenomenon we have 
also analyzed Eq. (18), to see if there is any possibility of
multiple sonic points. But we did not find multiple sonic points.
This limits the scope of formation of
radiative-hydrodynamic shocks in winds, although in time dependent
rotating flow, shocks can still form. \\

Jets originate from post-shock accretion flow close to
the compact object. And hence
outflows should be very hot. We have seen so far that for sufficiently
hot flow radiative acceleration has a very limited effect. In fact, very close
to the compact object cooling processes may be important. 
In the entire analysis so far, we ignored cooling processes.
We know that bremsstrahlung cooling (Chattopadhyay and Chakrabarti, 2000)
is
a very inefficient process. Therefore in the following analysis we include
the synchrotron loss due to stochastic magnetic field. 
We assume an equipartition magnetic field (B). The
synchrotron loss (erg $g^{-1}s^{-1}$) is given by Shapiro and Teukolsky,
(1983),
$$
S_L=\frac{16}{3}\frac{e^2}{c}(\frac{eB}{m_ec})^2(\frac{{k_B}{T_e}}{m_ec^2})^2
\frac{1}{m_p} ,
\eqno{(25)}
$$

\noindent where, $e$ and $m_e$ are the electron charge and its mass.
We include this cooling term in RHS of Eq. (4). Following
similar method as explained in \S 2, we get the sonic point condition.
The first condition is the same as Eq. (19) but the second condition
is given by:
$$
[1+\frac{({\gamma}-1){S_0}}{2(z_c-1)z_c}]{a_c}^{2}-z_c{\cal E}_za_c
+\frac{z_c}{2} [{\cal D}_{z_c}-\frac{1}{2(z_c-1)^{2}}]=0,
\eqno{(26)}
$$
\noindent where,
$S_0=(32{e^4}{{\mu}^2}{\pi}1.4{\times}{10^{17}}{\it f})/(3{{m_e}^3}
{c^3}{{\gamma}^2}{M_{\odot}}G{\Theta}_{out})$, where ${\it f}$
is the outflow rate in units of ${\dot M}_{Edd}$. \\

\begin {figure}
\vbox{
\vskip 0.0cm
\hskip -0.0cm
\centerline{
\psfig{figure=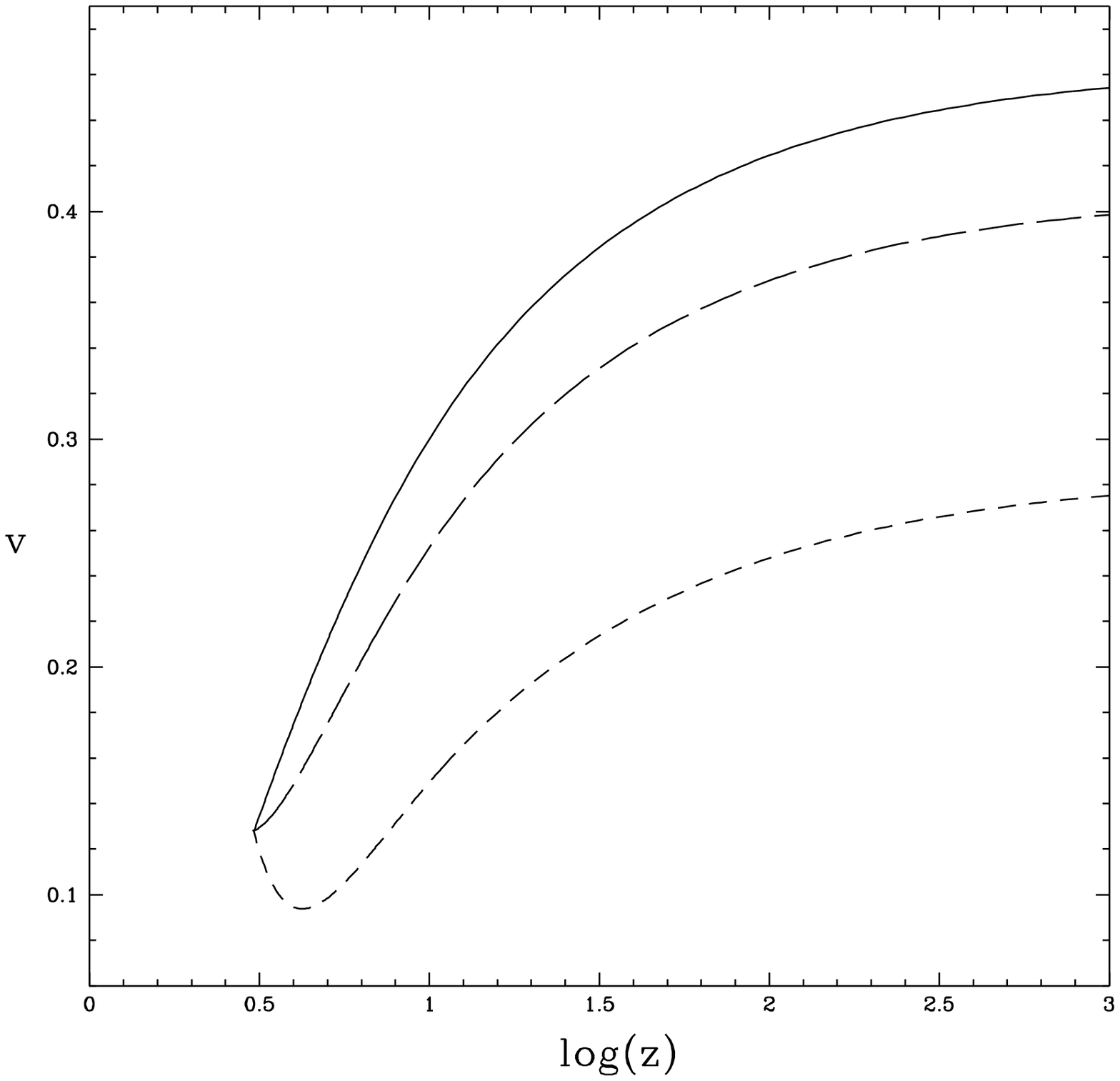,height=14truecm,width=14truecm}}}
\end{figure}
\begin{figure}
\vspace{-0.5cm}
\caption[] {Solution of outflows. Initial parameter
$v_{in}=0.13c$ at $z_{in}=3.05r_g$. RAMOD corresponds to ${\dot M}_{acc}
=6{\dot M}_{Edd}$. Variation of outflows without cooling (solid)
is contrasted with outflows with cooling for ${\dot M}=0.01{\dot M}_{Edd}$
(short-dashed) and ${\dot M}=0.005{\dot M}_{Edd}$ (long dashed). }
\end{figure}

Even in this case 
there is only one positive real root for
sonic point outside the horizon. Hence once again the possibility of finding
a steady shock in the transonic outflow is absent. 
Figure 6, compares the variation of
bulk velocity without (solid) cooling and with cooling for
${\dot M}=0.01{\dot M}_{Edd}$ (short-dashed) and ${\dot M}=0.005{\dot M}
_{Edd}$ (long dashed).
RAMOD corresponds to ${\dot M}_{acc}=6{\dot M}_{Edd}$.
Initial parameters are $z_{in}=3.05$ and $v_{in}=0.13$.
We can see that synchrotron  cooling produces less energetic outflow. However,
synchrotron loss rate (per unit volume) depends on the number density of electrons and
also the magnetic field strength. Number density of electrons
depends on the mass outflow rate (${\dot M}$). As we have no theoretical
handle on the magnetic field, we have assumed an equipartition magnetic field ($B$),
which depends on the density of matter and hence on ${\dot M}$. Thus tuning
${\dot M}$ would also tune $B$. We reduce ${\dot M}$ to $0.005{\dot M}_{Edd}$ 
(long-dashed) and find that compared to the
solution which includes cooling (long-dashed) with
higher ${\dot M}$ results in outflows which are more
energetic. If one had a better understanding of magnetic fields
around accreting black holes, then synchrotron cooling phenomenon would be worth pursuing.
This is just to show cooling can be very important close to the
black hole. 

\section{Discussion and Concluding Remarks}

It has been pointed out in \S 1, that outflows around compact objects
and especially black holes have to originate from the inner part of the
accreting matter. This should, in-fact be a region which is also the source of
hard photons. As a photon interacts
with an electron, Eq. (1) tells us that there will be exchange of
energy between them and therefore momentum. If the thermal energy of the
electron is less than the photon energy it will gain energy from the
photon. But this does not ensure enhancement of bulk kinetic energy
of the flow. Energy gained may not enhance 
its kinetic energy. Intense radiation in the vicinity
of the black hole would create sufficient radiative pressure on the
flow to slow it down due to radiation drag. So even if one
considers only cold plasma even then radiative acceleration of outflows
remains a tricky issue which we have tried to resolve in the present paper. 

We have witnessed that as one increases radiative momentum deposition force, one gets more and more
energetic outflows, but there is no linear relation between 
velocity at large distance and the radiative force. 
This is because of radiation drag effect.
Higher radiative force gives rise to a higher energy density. In principle,
higher radiative force means more acceleration  resulting in higher
velocity, but as radiation drag term is proportional to both $v$ and
energy density, therefore the deceleration starts to be important too.
Equating the first term and the second term in RHS of Eq. (2),
we get the expression of equilibrium speed $v_{eq}$.
If one considered the disc to be an infinite flat radiator, as
Icke (1989) did, at large distances, the 
contribution to the radiative energy density remains higher compared to the
flux along the axis of symmetry. This is not true when the
`radiatior' has a special geometry as in our case. Here,
within the {\it funnel} of the inner torus of the  {\it finite} accretion disc, radiation 
received at any point on the axis is significantly larger than the 
flux along the axis. As one goes further and further away 
from the inner torus, the matter will increasingly
see it as a point source. Hence the energy density will be that due to a
point source. The efficiency of `flux-focussing' is thus better. As a result,
we get marginally higher equilibrium velocity (at a large distance) than that 
obtained for a thin accretion disc. 
Icke (1989) has also computed magic speed for the thick disc where the 
temperature of the inner surface of the thick disc decreases with
the height. He showed that this temperature variation of the driving
surface significantly influence the value of magic speed.
In our case the radiation coming out of the inner torus are
inverse Comptonized photons and are not thermalized with the post-shock
torus. Moreover, the CENBOL surface is an isothermal one. But the frequency 
averaged intensity will depend on the temperature gradient at the surface
of the CENBOL. With the particular case we have taken, the temperature gradient
is seen not to vary appreciably except very close to the black hole.
Hence for simplicity we have taken the intensity of radiation to be uniform.

Recently Fukue and his collaborators (1999-2001) has systematically
studied the radiative acceleration and collimation of jets coming out
from a disc which is a combination of inner ADAF region (non luminous)
and outer slim disc (luminous). First of all, the ADAF solutions are for
very low accretion rate (${\dot M}_{acc} {\lsim} 10^{-5}{\dot M}_{Edd}$), 
so the outflow rate would be even smaller.
The second fact is that the radiative collimation they have got 
is only because of the inner non-luminous region which 
does not contribute to the
radial ($r, {\phi}, z$ system) flux. The situation is different in our case.
The hard radiation is coming from the inner post-shock region of the
disc while the soft radiation is coming from the outer region of disc. 
In this paper we have only considered the radiations coming from
the post-shock region. Therefore we are working in a different regime.

We ignored bremsstrahlung cooling because it is a very inefficient
cooling mechanism and is not likely to change the physics of outflows,
qualitatively. We also have ignored cooling due to inverse-Comptonization. 
As we initially were interested to
look at acceleration phenomenon of the radiation we had to look for
cold plasma. Thus the equations of motion presented in \S 2. are not
suitable to study the effect of inverse Comptonization.
In future we will work with the most general form of radiative transfer
equation in curved space time, which will include all the processes.

While calculating the specific intensity of radiation (in \S 2.3) coming out
from the inner torus of the accretion disc, we have assumed that the 
radiation coming out is proportional to the gravitational energy release
of in-falling matter. The conversion ratio (${\eta}$) used, was that due 
to last stable
orbit around a Schwarzschild black hole. As the Keplerian disc only 
extends upto the shock location (i.e., 
$R_s$ in \S 2.4) hence ${\eta}$ will be less than $0.06$, around $0.04$.
Hence there is a little bit of over estimation in the calculation
of the terminal velocity, though
this is not going to change the result qualitatively.

Our conclusion in the present paper is that the
radiation momentum deposition does accelerate matter to
produce transonic outflows.
Radiative acceleration increases with the
accretion rate. Achievement of very high 
velocity is restricted because of the radiation drag term.  
Radiation drag term is the deceleration term and it 
depends on both the velocity of the flow and the energy density of radiation.
As a result above some speed ($v_{eq}$) radiative deceleration sets in.
Within the funnel of the inner post-shock torus of the accretion disc,
$v_{eq}$ depends on distance but at large distance from it the steady value
attained is $0.5$ i.e. fifty percent of the velocity of light.
It is to be noted that $v_{eq}$ itself is not the terminal
velocity $v_\infty$ -- it becomes the terminal velocity for the flow
having an infinitesimally small initial energy $E_{in}$ [see, Eq. 23(c)]. If 
$E_{in}>0$, it can be easily shown that  $v_\infty^2 < v_{eq}^2 + 2 E_{in}$. 

\noindent Radiative-hydrodynamics changes the critical point condition,
but fails to generate multiple real critical points
of the equation thus eliminating the possibility
of getting steady shocks in outflows. This may be related with the assumptions
enforced, namely, due to the consideration of cold plasma in
a {\it non-rotating} flow. Radiative momentum deposition increases the 
specific energy of the outflow to the extent of driving bound outflowing 
matter to infinity as transonic outflow. 
Synchrotron cooling results in less energetic outflow, but
this effect can increase with larger outflow rate. A proper theoretical
handle on the structure of magnetic field will dramatically affect
the physics of outflows. 

The results presented here are calculated bearing a bipolar
outflow in mind. It will be noticeable that, the parameters, like
the accretion and outflow rates,
used here are on the higher side. This was done to
study the effects in more extreme conditions. 
Essential features observed here will remain
unaffected for the lower values of the parameters as well.

{}
\end{document}